\documentclass[aps,pre,preprint,groupedaddress,showpacs]{revtex4}
\usepackage[dvips]{graphicx}

\begin{document}

%\preprint{}

\title{Statistics for transition of a plasma turbulence with 
multiple characteristic scales}

\author{Mitsuhiro Kawasaki}
\email{mituhiro@riam.kyushu-u.ac.jp}
\homepage{http://www.riam.kyushu-u.ac.jp/sanny/activity/member/mituhiro/mituhiro-e.html}

\author{Sanae-I. Itoh}
\affiliation{Research Institute for Applied Mechanics, Kyushu University, Kasuga 816-8580, Japan}

\author{Kimitaka Itoh}
\email{itoh@nifs.ac.jp}
\affiliation{National Institute for Fusion Science, Toki 509-5292, Japan}

\date{\today}

\begin{abstract}
Subcritical transition of an inhomogeneous plasma where turbulences with 
different characteristic space-time scales coexist is analyzed with methods of 
statistical physics of turbulences. 
We derived the development equations of the probability density function 
(PDF) of the spectrum amplitudes of the fluctuating electro-static potential. 
By numerically solving the equations, the steady state PDFs were obtained. 
Although the subcritical transition is observed when the turbulent 
fluctuations are ignored, the PDF shows that the transition 
is smeared out by the turbulent fluctuations. It means that the approximation 
ignoring the turbulent fluctuations employed by traditional transition 
theories could overestimate the range where hysteresis is observed and 
statistical analyses are inevitably needed.
\end{abstract}

\pacs{52.55.Dy,52.55.-s,52.30.,47.27.-i}
%\keywords{plasma turbulence, submarginal turbulence, subcritical 
%bifurcation, probability density function, Langevin equation, Fokker-Planck 
%equation, current diffusive interchange mode, ion temperature gradient mode}

\maketitle

% body of paper here
\section{\label{section-introduction}Introduction}
Transition phenomena with sudden changes of states are observed in 
turbulent plasmas. Since these transition phenomena like the L-H transition 
play crucial roles for magnetic confinement of fusion plasmas, 
the transition associated with formation of transport barriers is one of the 
main subjects of high-temperature plasma physics.

Traditional theories for transition of high-temperature plasmas are 
formulated in terms of averaged physical quantities 
\cite{itoh1988, shaing1989}. 
Fluctuations around the averages are ignored and 
transition phenomena are described deterministically.

However, high-temperature plasmas are strongly non-linear systems with 
huge number of degrees of freedom and hence their behavior should be 
chaotic and unpredictable in a deterministic way. 
In fact, broad distribution of critical values of parameters where 
transitions occur and intermittent transport called ``avalanche phenomena'' 
are observed in recent experiments 
\cite{iter1994, politzer2000, yoshizawa2002}.
Occurrence of these behaviors, which cannot be described only by averaged 
quantities, are considered due to strong turbulent fluctuations. 
It is inevitable to describe turbulent plasmas in terms of probability or 
ensembles like numerical forecast of weather, since the magnitudes of 
fluctuations around averages are of the same order as that of the averages in 
turbulent states.

Therefore, transition takes place as a statistical process in the 
presence of stochastic noise sources induced by turbulence interactions. 
As a generic feature, transition is expected to occur with a finite 
probability when a control parameter approaches the critical value.

Statistical theories for plasma turbulence have been developed and the 
framework to calculate the probability density function (PDF), the transition 
probability etc. has been made \cite{bowman1993, krommes1996, krommes1999, 
krommes2002, itoh1999, itoh2000, itoh2002, yoshizawa2002}. 
In the statistical theories, 
the time-development of the system is described by 
a set of differential equations with random forces, called the 
``Langevin equations''. All the information on the statistical properties of 
the system is obtained by solving the Langevin equation.

The framework has been applied to cases where 
only one turbulent mode is excited and the turbulence is characterized by 
one space-time scale \cite{bowman1993, krommes1996, krommes1999, krommes2002, 
itoh2000, itoh2002-2, kawasaki2002, kawasaki2002-2}. 

However, it is well known that there are many kinds of turbulent fluctuations 
in high-temperature plasmas and that different characteristic length scales 
coexist. The importance of interactions between modes with different scale 
lengths has recently been recognized. 
For instance, the dynamics of the meso-scale 
structure of the radial electric field 
\cite{biglari1990, itoh1999b, terry2000} is known to cause 
variation in the dynamics of microscopic fluctuations like in 
the electric filed domain interface \cite{itoh1991,diamond1997}, 
zonal flow \cite{smolyakov1999} and streamer \cite{drake1995}. 
Coexistence of multiple scale turbulence has also been investigated by use of 
the direct numerical simulations \cite{yagi2002, kishimoto2002}. 
Statistical theory on zonal flow dynamics \cite{krommes2000} and that of the 
L-H transition theory has been developed \cite{itoh2002-2}.

In the present paper, we apply the statistical theoretic algorithm to a model 
of high-temperature plasma where two characteristic scales coexist. 
One is the current diffusive interchange mode (CDIM) micro turbulence 
\cite{itoh1999b}, 
whose characteristic length scale is of the order of 
the collisionless skin depth $\delta = c/\omega_p$. 
The other is the ion-temperature gradient (ITG) mode turbulence, 
whose characteristic wave length is of the order of the ion gyroradius 
$\rho_i$, as an example of the drift wave fluctuations considered to 
dictate a considerable part of the turbulent transport \cite{horton1999}. 
Hereafter, we call these two modes ``the high wave number mode'' and 
``the low wave number mode'' 
respectively for its simplicity. We assume that the condition 
$\rho_i \gg \delta$ holds. Both turbulences are considered to cause 
the anomalous transport and hence the coexistence of the high wave number mode and 
the low wave number mode turbulences and their interplay should be taken into account. 

It is known that the subcritical transition occurs in this system, 
when the pressure-gradient and the radial electric field are changed 
\cite{itoh2001}. However, the turbulent fluctuations are ignored in the 
analysis. In the present paper, with the statistical theory, 
we analyze the effect of the turbulent 
fluctuations on stochastic properties of the transition and 
show that the fluctuation changes the phase structure of the system 
completely. More precisely, we show that the transition is smeared out by 
noises, i.e., the physical quantities changes gradually without clear 
transition. 

The present paper is organized as follows; the statistical theory and the 
model are formulated in Sec.\ \ref{section-model}. The results of 
deterministic analyses including occurrence of the subcritical bifurcation 
are summarized in Sec.\ \ref{section-bifurcation}. The Section\ 
\ref{section-stochastic}, i.e., the main part of this paper, 
presents the statistical properties of the system. 
Summary and discussions are given in Sec.\ \ref{section-summary}.

\section{\label{section-model}Theoretical framework and the model}

In this section, we briefly review the theoretical framework and the model 
of turbulent plasmas where two different characteristic scales coexist 
\cite{itoh2001}.

The starting point of the theory is the reduced MHD for the three fields: 
the electro-static potential, the current and the pressure. 
The Langevin equation that the statistical theory is based upon 
is derived as one of model equations which reproduce 
the two-time correlation functions and the response 
functions obtained by the renormalization perturbation theory 
(the direct-interaction approximation) for the reduced MHD \cite{krommes2002}. 

The Langevin equation describes time-development of two variables 
characterizing the system. 
One is the spectrum amplitude of the electro-static potential for 
the characteristic wave number $k_h$ of the high wave number mode,  
$x_h \equiv k_h^2 \langle \phi^*_{k_h} \phi_{k_h} \rangle/D_h^2$, and 
the other is that for the low wave number mode, 
$x_l \equiv k_l^2 \langle \phi^*_{k_l} \phi_{k_l} \rangle/D_l^2$, 
whose characteristic wave number is $k_l$. 
Here, $D_h$ and $D_l$ denote the renormalized transport coefficients 
for the high wave number mode and the low wave number mode respectively when there is no interactions 
between two modes. 
The characteristic time constants for the two modes are defined as 
\begin{equation}
\omega_l \equiv k_l^2 D_l, \omega_h \equiv k_h^2 D_h.
\label{time-constant}
\end{equation}
See \cite{itoh2001} for the details of the notation.

The Langevin equation gives the time-development of these two variables as 
\begin{eqnarray}
\frac{dx_l}{d(\omega_l t)} + \frac{1}{2}\left(\frac{\sqrt{x_h}}{r}+\sqrt{\frac{x_h}{r^2}+4 x_l}-2\right) x_l 
+\frac{1}{2}\left(\frac{\sqrt{x_h}}{r}+\sqrt{\frac{x_h}{r^2}+4 x_l} \right) x_l  w_l = 0, 
\label{Langevin-eq-l}
\\
\frac{dx_h}{d(\omega_h t)}+\left[ \sqrt{x_h}-\sqrt{1+\frac{\sqrt{x_l}}{\sqrt{x_h}/r+\sqrt{x_h/r^2+4 x_l}}}/(1+p r^2 x_l) \right] x_h 
+ x_h^{3/2} w_h+\sqrt{\epsilon} w_t = 0.
\label{Langevin-eq-h}
\end{eqnarray}
Nonlinear drags and magnitudes of the noises have been evaluated in 
\cite{itoh2001}.

In these equations, the nonlinear terms of the reduced MHD are divided into 
two parts: One part is coherent with the test field and is renormalized into 
the deterministic terms, i.e., the second terms. 
The other is incoherent and is modeled 
by random forces $w_h, w_l$. Another random force $w_t$ denotes the 
thermal noise and the magnitude of the noise, $\epsilon$, is a small quantity 
compared to the magnitudes of other noises. 
Technically, introduction of $\epsilon$ is needed in order to exclude 
singularity at $x_h = 0$ of the PDF. 
%The value of $\epsilon$ is determined from the temperature.

Since $w_h, w_l$ and $w_t$ are forces which fluctuates randomly in time, 
the Langevin equation describes the stochastic time-development of the 
turbulent fluctuation of the system. 
For simplicity, we assume that the random forces are Gaussian and white;
\begin{equation}
\langle w(t) w(t') \rangle = \delta(t-t').
\end{equation}

This system is sustained by the space inhomogeneities, 
the curvature of the magnetic field $\Omega'$, 
the pressure gradient $dP_0/dx$ and the gradient of the radial electric field 
$dE_r/dr$. Here, the shear of the magnetic field is given as the slab 
configuration: 
$\mathbf{B}=(0,B_0 s x, B_0)$ where $B_0(x) = \mbox{const}\times 
(1+\Omega' x+\cdots)$. The pressure is assumed to change in $x-$direction.
These driving forces are characterized by the parameter $r$ 
and $p$ as $r \equiv D_l/D_h \propto (\Omega' dP_0/dx)^{-1/2}, 
p \equiv D_h^2/I_{eff}^{h\leftarrow l} \propto (dE_r/dr)^{-6} \Omega' 
dP_0/dx$. Here, $I_{eff}^{h\leftarrow l}$ denotes the critical strength 
of the nonlinear interactions between the low and high wave number modes. 
The pressure gradient controls the growth rate of the low wave number mode and 
$\Omega' dP_0/dx$ excites both the high and low wave number mode turbulences. 
The gradient of the radial electric field suppresses turbulences 
\cite{biglari1990, itoh1999b, terry2000}.

We assume that the relation $\rho_i \gg \delta$ holds and hence the 
characteristic length-scales for the two modes are widely separated as 
\begin{equation}
k_h \gg k_l.
\label{scale-separation}
\end{equation} 
The mutual interactions between the low and high wave number modes are 
asymmetric, since the spatial structure of the low wave number mode is a large-scale 
inhomogeneity for the high wave number mode.
The assertion Eq.\ (\ref{scale-separation}) also means that the time-scales 
are widely separated since the time scales are given by 
Eq.\ (\ref{time-constant}); 
\begin{equation}
\omega_h \gg \omega_l.
\label{time-scale-separation}
\end{equation} 

By analyzing the Langevin equation, Eqs.\ (\ref{Langevin-eq-l}, 
\ref{Langevin-eq-h}), a number of 
statistical properties of turbulent plasmas can be derived. For example, 
the analytical formulae of the rate of change of states of plasmas, the 
transition rates, were derived. Furthermore, since the renormalized transport 
coefficients and the random forces have the same origin, i.e., 
nonlinear interactions in MHD turbulence, relations between the 
fluctuation levels of turbulence and the transport coefficients like the 
viscosity and the diffusivity were derived.

\section{\label{section-bifurcation}Bifurcation without random forces}

With the theoretical framework briefly described in the previous section, 
we analyze the model of the inhomogeneous plasma with two characteristic 
scales.

At first, we show the steady state solutions when random forces are ignored, 
in order to compare to the results with the random forces obtained later. 
The steady state solutions are obtained by solving the set of 
nonlinear equations when the random forces are turned off:
\begin{eqnarray}
\left( \frac{\sqrt{x_h}}{r}+\sqrt{\frac{x_h}{r^2}+4 x_l}-2 \right) x_l = 0, 
\label{bifurcation-l}
\\
\left[ \sqrt{x_h}-\sqrt{1+\frac{\sqrt{x_l}}{\sqrt{x_h}/r+\sqrt{x_h/r^2+4 x_l}}}/(1+p r^2 x_l)\right] x_h = 0.
\label{bifurcation-h}
\end{eqnarray}

\begin{figure}
\includegraphics[width=10cm, keepaspectratio]{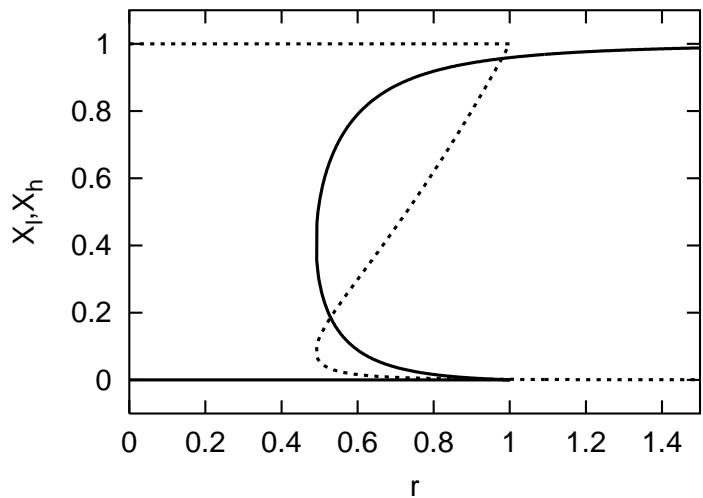}
\caption{\label{figure-bifurcation}The $r$-dependence of the steady state 
solutions when no random force. The parameter $p$ is fixed as $p=30$. 
The solid line represents the low wave number mode $x_l$ 
and the dotted line represents the high wave number mode $x_h$. 
It is seen that the subcritical bifurcation occurs.}
\end{figure}

Figure\ \ref{figure-bifurcation} shows the $r$-dependence when $p=30$. 
When $r \le 0.49\ldots$, the low wave number mode turbulence is suppressed. 
As $r$ is increased, the system experiences the subcritical transition to 
the state where the low wave number mode turbulence is excited. 
When $0.49\ldots < r < 1$, there are two stable solutions and it means that 
the system is bi-stable. 
From the deterministic point of view, the transition is expected to occur 
only at the ridge point and the bifurcation point. 
The qualitative behavior does not depend on the value of $p$ as far as 
$p > 1.9 \ldots$.

\begin{figure}
\includegraphics[width=10cm, keepaspectratio]{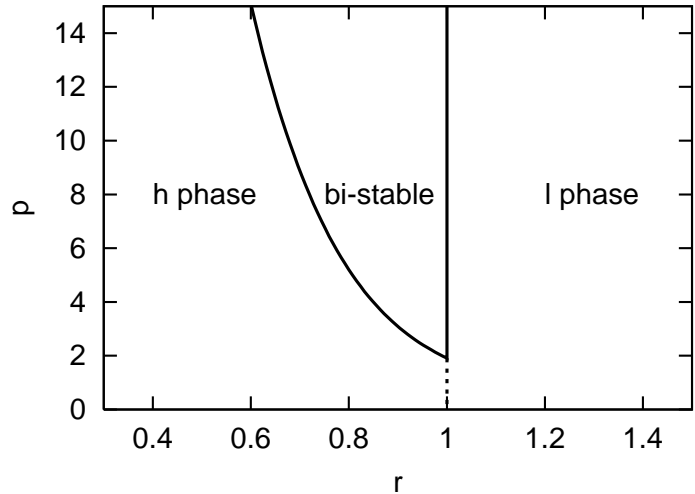}
\caption{\label{p-r-diagram}The phase diagram when the random forces are 
turned off. The subcritical bifurcation is observed when 
$p > 1.9\ldots$ and the boundaries of the bi-stable phase are represented 
with the solid line. When $p \leq 1.9\ldots$, the bifurcation is 
supercritical and critical values are plotted with the dotted line. 
The region indicated as ``h phase'' is the state where 
the high wave number mode is excited and the region ``l phase'' is the state where 
the low wave number mode is excited.}
\end{figure}

The phase diagram is shown in Fig.\ \ref{p-r-diagram}. 
The subcritical bifurcation is observed when the value of the parameter $p$ is 
larger than $1.9\ldots$. On the other hand, 
when $p \leq 1.9\ldots$, the bifurcation is supercritical. 

\section{\label{section-stochastic}The stochastic properties}

In the rest of the present paper, we analyze the stochastic properties of 
the model, Eqs.\ (\ref{Langevin-eq-l}, \ref{Langevin-eq-h}), 
to investigate the effect of 
the turbulent fluctuations, i.e., the random forces.

\subsection{The adiabatic approximation and the Fokker-Planck equation}

At first, we approximate the Langevin equation Eqs.\ (\ref{Langevin-eq-l}, 
\ref{Langevin-eq-h}) with making use of the time scale separation 
between the low and high wave number mode turbulences. 

The scale separation, Eq.\ (\ref{time-scale-separation}), means that 
the high wave number mode $x_h$ quickly relaxes to the steady state determined by the 
value of the low wave number mode variable $x_l$ which is fixed at the value $x_l=x_l(t)$ 
at the time.

We analyze the steady state of the high wave number mode $x_h$ when $x_l$ is fixed.
A state of a stochastic system is described by the probability that 
a state variable takes a certain value. The time-development of the 
probability density function (PDF) $P(x_h,t)$ of $x_h$ is determined by 
Kramers-Moyal expansion applied to the Langevin equation of $x_h$, 
Eq.\ (\ref{Langevin-eq-h}). The Kramers-Moyal expansion is given as 
\begin{equation}
\frac{\partial P(x_h,t)}{\partial t} = -\frac{\partial}{\partial x_h}
C_1(x_h)P(x_h,t)+\frac{1}{2}\frac{\partial^2}{\partial x_h^2} C_2(x_h) 
P(x_h,t).
\end{equation}
The expansion can be truncated at the second order, since the 
random forces in the Langevin equation are assumed to be Gaussian.
Here, the coefficient $C_n(x_h)$ is given by 
\begin{equation}
C_n(x_h) \equiv \lim_{\triangle t \rightarrow 0} \frac{1}{\triangle t} 
\left\langle [x_h(t+\triangle t)-x_h(t)]^n\right\rangle,
\end{equation}
where $\langle\cdot\rangle$ denotes the average over the all realizations 
of the random forces and the average is taken under the condition 
$x_h(t) = x_h$. The resulting equation of motion of the 
probability density function is called the Fokker-Planck equation and is 
written as 
\begin{eqnarray}
\frac{\partial P(x_h,t)}{\partial (\omega_h t)} = 
\frac{\partial}{\partial x_h} \left[ \sqrt{x_h}-
\sqrt{1+\frac{\sqrt{x_l}}{\sqrt{x_h}/r+
\sqrt{x_h/r^2+4 x_l}}}/(1+p r^2 x_l)\right] \nonumber \\ 
\times x_h P(x_h,t)+\frac{1}{2} \frac{{\partial}^2}
{\partial x_h^2}(x_h^3+\epsilon)P(x_h,t).
\label{fk-h}
\end{eqnarray}

\begin{figure}
\includegraphics[width=10cm, keepaspectratio]{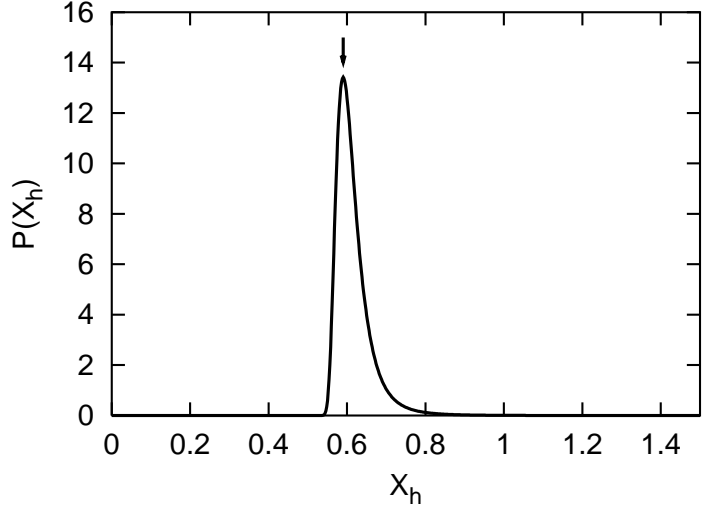}
\caption{\label{pdf-xf}The steady state PDF $P(x_h)$ 
when $x_l$ is fixed at $x_l=1$. The peak of the PDF is relatively narrow and 
the adiabatic approximation that $x_h$ is replaced with the peak value of the 
PDF, $x_h^*(x_l)$, is justified. The arrow indicates the location of the peak.}
\end{figure}

The steady state solution of Eq.\ (\ref{fk-h}) when $x_l$ is fixed 
is shown in Fig.\ \ref{pdf-xf}. It is seen that the peak of the PDF is 
relatively narrow and it means that the high wave number mode turbulence spends most 
of time at the peak value $x_h^*$. Hence, we can say that 
it is good approximation to replace $x_h$ in the Langevin equation of the 
low wave number mode with the peak value $x_h^*$. 

The equation to determine the location of the peak $x_h^*$ is given 
by the condition $d P(x_h^*)/d x_h^* = 0$ 
%from the Fokker-Planck equation Eq\ (\ref{fk-h}) 
as 
\begin{equation}
\sqrt{1+\frac{\sqrt{x_l}}{\sqrt{x_h^*}/r+\sqrt{x_h^*/r^2+4 x_l}}}/
(1+p r^2 x_l)-\frac{3}{2} x_h^*-\sqrt{x_h^*}= 0.
\label{peak-h}
\end{equation} 
It is important to note that Eq.\ (\ref{peak-h}) is essentially different from 
Eq.\ (\ref{bifurcation-h}) in existence of the second term of 
Eq.\ (\ref{peak-h}), which comes from the random force $x_h^{3/2} w_h$ of 
Eq.\ (\ref{Langevin-eq-h}). It implies that 
the random forces change the steady state of the high wave number mode.

Consequently, the adiabatically approximated Langevin equation for the low 
wave number mode is given by Eq.\ (\ref{Langevin-eq-l}) where $x_h$ is 
replaced with 
$x_h^*$ determined by Eq.\ (\ref{peak-h}). The reduced Langevin equation is 
written as 
\begin{equation}
\frac{dx_l}{d(\omega_l t)} + \frac{1}{2}\left(\frac{\sqrt{x_h^*}}{r}
+\sqrt{\frac{x_h^*}{r^2}+4 x_l}-2\right) x_l 
+\frac{1}{2}\left(\frac{\sqrt{x_h^*}}{r}+\sqrt{\frac{
x_h^*}{r^2}+4 x_l} \right) x_l w_l = 0. 
\label{reduced-Langevin}
\end{equation}
The corresponding Fokker-Planck equation, which determines the 
time-development of the PDF of $x_l$, $P(x_l,t)$, is given by 
\begin{eqnarray}
\frac{\partial P(x_l,t)}{\partial (\omega_l t)} & = & \frac{1}{2}
\frac{\partial}{\partial x_l} \left(\frac{\sqrt{x_h^*}}{r}
+\sqrt{\frac{x_h^*}{r^2}+4 x_l}-2 \right) x_l P(x_l,t) \nonumber \\ && 
+\frac{1}{2}\frac{{\partial}^2} 
{\partial x_l^2} \left[ \frac{1}{4} \left(\frac{\sqrt{x_h^*}}{r}
+\sqrt{\frac{x_h^*}{r^2}+4 x_l} \right)^2 x_l^2 \right] P(x_l,t).
\label{fk-l}
\end{eqnarray}

\subsection{\label{section-results}The probability density functions and 
the effect of the random forces}

In this subsection, we investigate properties of the steady state PDF of 
the low wave number mode, $P(x_l)$, by numerically solving Eq.\ (\ref{fk-l}). 
Although we will show figures when the parameter $p$ is fixed at $p=30$, 
qualitative behavior is the same as that for other $p$ values larger than 
$1.9 \ldots$.

\begin{figure}
\includegraphics[width=10cm, keepaspectratio]{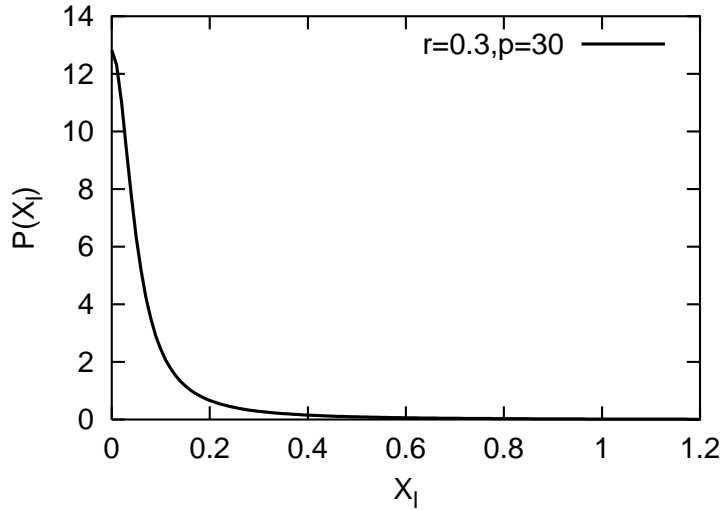}
\caption{\label{pdf-h-phase}The steady state PDF $P(x_l)$ in the small $r$ 
region. It shows that the probability that the low wave number mode is suppressed is large 
and the result is compatible with that of the deterministic analyses given 
in Fig.\ \ref{figure-bifurcation}.}
\end{figure}

At first, we show in Fig.\ \ref{pdf-h-phase} the steady state PDF in the small 
$r$ region. 
In this region, we have seen that the low wave number mode is suppressed when 
the turbulent fluctuations are ignored. Figure\ \ref{pdf-h-phase} shows that 
the probability that the low wave number mode is quiet is large and hence 
the analysis ignoring random forces is a good approximation in this region.

\begin{figure}
\includegraphics[width=10cm, keepaspectratio]{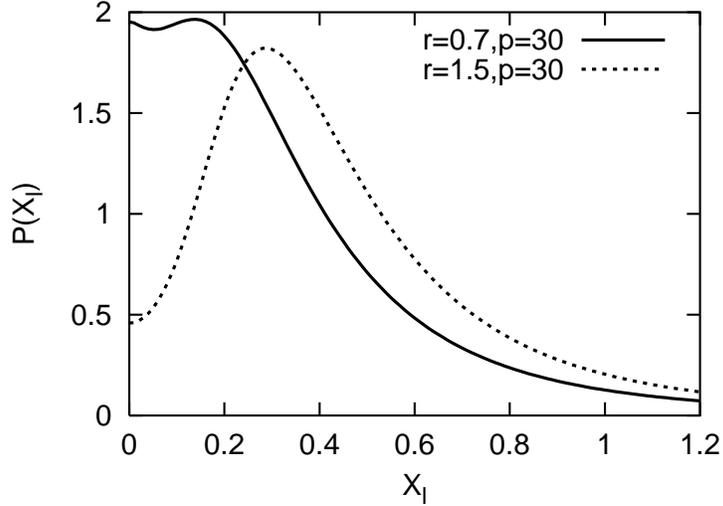}
\caption{\label{ss-pdf-l-r}The steady state PDFs of $x_l$ when 
$0.49\ldots < r$. The solid line represents the PDF for $r=0.7$ and 
the dotted line represents the PDF for $r=1.5$. The wide peak which 
consists of two small unresolved peaks means that the subcritical transition 
cannot be observed due to the turbulent fluctuations.}
\end{figure}

However, as the value of the parameter $r$ is increased, the characteristics 
change completely. The steady state PDF when $r > 0.49\ldots$ is shown in 
Fig.\ \ref{ss-pdf-l-r}.
In the region $0.49\ldots < r < 1$, we have seen that the subcritical 
transition occurs and the system is bi-stable 
if the random forces are ignored. Although there are two peaks in the 
PDF which are compatible with the previous result without random forces, 
the valley between the peaks is too shallow to identify the two states. 
In other words, even if one observes the time-series of $x_l(t)$, 
the value of $x_l(t)$ strolls around the two peaks without a sudden change. 
It means that the bi-stability of the system, i.e., the 
subcritical transition, is smeared out by 
the turbulent fluctuations.

The steady state PDF for $r > 1$ is also shown in Fig.\ \ref{ss-pdf-l-r}. 
It is seen that the PDF has a single peak. 
The single peak is compatible with the result of the deterministic 
analysis, where only one state exists. However, the peak is wide and 
hence $x_l$ fluctuates widely around the location of the peak. 
The variance of the fluctuation is as large as the peak value and it means 
that the fluctuation cannot be ignored. 
Furthermore, the peak value $0.29\ldots$ obtained when $r=1.5, p=30$ is 
different from the root of Eqs.\ (\ref{bifurcation-l}, \ref{bifurcation-h}), 
which is given as $x_l \simeq 0.98$.

\begin{figure}
\includegraphics[width=10cm, keepaspectratio]{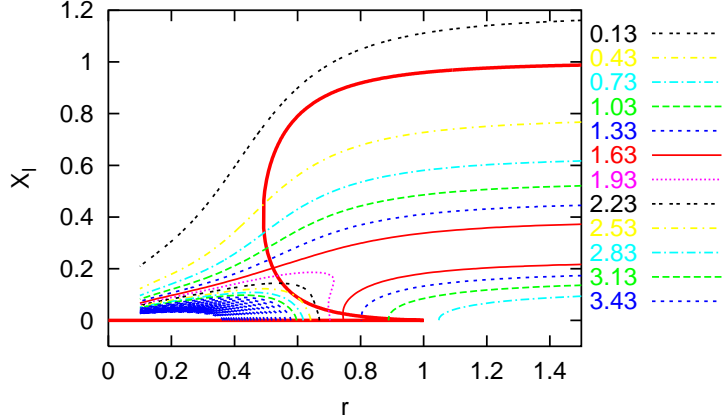}
\caption{\label{contour-ss-pdf}The contour plot of the $r$-dependence of 
the PDFs. The thin lines represent the contour lines. 
All the contours larger than $3.43$ are displayed with the same dotted lines. 
The bifurcation diagram represented with the bold solid line is added for 
comparison. The location of the region with large probability, 
mainly denoted with the thin solid lines, is shifted 
as the value of the parameter $r$ is changed. However, there is no 
singularity on that.}
\end{figure}

In order to see these consequences explicitly, 
in Fig.\ \ref{contour-ss-pdf}, we show the comparison of 
the contour plot of the PDF and the bifurcation diagram obtained by the 
analysis ignoring the random forces.  
It is seen that the location of the region which has large probability 
and mainly denoted with the thin solid lines in Fig.\ \ref{contour-ss-pdf} 
is shifted gradually as the value of the parameter $r$ is changed. 
Furthermore, the location of the peak of the PDF is far from the result when 
the random forces are ignored when $r \geq 0.49\ldots$.

\begin{figure}
\includegraphics[width=10cm, keepaspectratio]{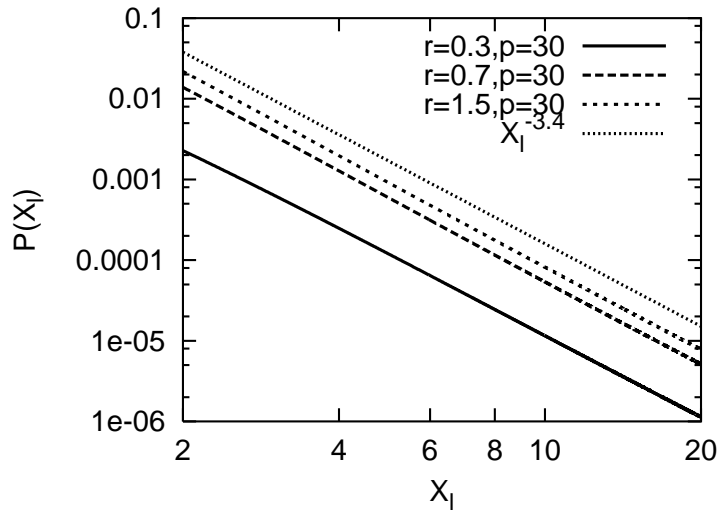}
\caption{\label{ss-pdf-log}The log-log plot of the tails of the steady state 
PDFs for different values of $r$. 
All the tails are well-approximated by the same power-law.}
\end{figure}

\begin{figure}
\includegraphics[width=10cm, keepaspectratio]{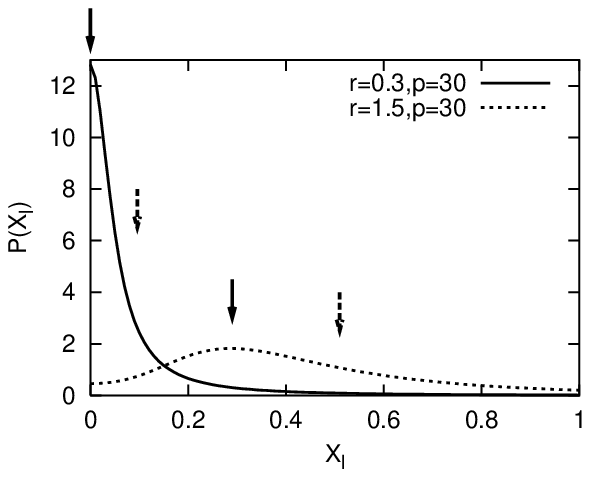}
\caption{\label{ave-mprob}Comparison of the averages with the most probable 
values. The averages are indicated with the dashed arrows and the 
most probable values (the peaks of the PDFs) are indicated with the solid 
arrows. The difference due to the long tail shown in Fig.\ \ref{ss-pdf-log} 
means that the average is not equal to the value which $x_l$ takes with the 
largest probability.}
\end{figure}

Next, in order to investigate the meaning of the average in such a 
turbulent system, we analyze the tail of the steady state PDF $P(x_l)$ 
shown in Fig.\ \ref{ss-pdf-log}.
It is shown that the tail of the PDF is well-approximated by the power-law 
and hence the probability is distributed broadly over large values of 
the fluctuation. The exponent is about $-0.34$ for the case shown in 
Fig.\ \ref{ss-pdf-log}. 
It means that the average, i.e., ``the center of mass'' of 
the PDF, is shifted in the large $x_l$ direction with the long tail and hence 
the average is not equal to the value that $x_l$ takes with large probability. 
The values that $x_l$ takes with large probability are characterized by the 
peaks of the PDF and are called the most probable values. 
%On the other hand, the value that $x_l$ takes with the largest 
%probability, which is called the most probable value, 
%is given by the peak of the PDF. 
The difference between the average and 
the most probable value is shown in Fig.\ \ref{ave-mprob}.

\begin{figure}
\includegraphics[width=10cm, keepaspectratio]{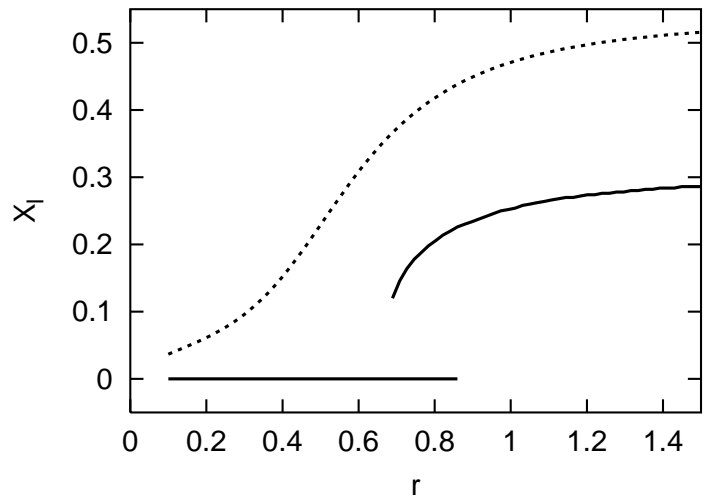}
\caption{\label{m-prob-average} The $r-$dependence of the most probable 
values (solid lines). The average is also plotted for comparison 
(a dotted line). Although relatively steep variation of the average is 
observed, the hysteresis can be seen only for the most probable values.}
\end{figure}

Furthermore, hysteresis cannot be captured by observing the average, since 
the quantity is single-valued from its definition. On the other hand, 
the most probable values depict the hysteresis as shown in 
Fig.\ \ref{m-prob-average}. 
Although relatively steep variation of the average is observed, 
the hysteresis can be seen only for the most probable values.

\section{\label{section-summary}Summary and discussion}

Finally, we summarize our results and consider their implications. 
We applied the statistical theory to the model 
of the inhomogeneous turbulent plasma where two turbulences well-separated 
in those space-time scales coexist; 
the ``high wave number mode'' turbulence (the CDIM micro turbulence) 
and the ``low wave number mode'' turbulence 
(the ITG mode semi-micro turbulence). 
We derived the development equations of the PDFs of the spectrum amplitudes 
of the electro-static potential for the characteristic wave numbers. 
By numerically solving the adiabatically approximated Fokker-Planck equation, 
the steady state PDFs for the low wave number mode turbulence were obtained. 
Although the subcritical bifurcation is observed when the turbulent 
fluctuations are ignored, the shape of the PDF shows that the transition 
is smeared out by the fluctuations. It means that the approximation 
ignoring the turbulent fluctuation like the traditional transition theories 
could overestimate the range of cusp catastrophe.

We also compared the average values with the most probable values and 
showed that these two characteristic values of stochastic nature 
are different due to the long power-law tail of the PDF. 
It means that the average does not mean the value which is expected to 
realize most probably and hence description of the state of the turbulent 
systems 
needs not only the average but also other statistical quantities like 
the most probable values and the variances.

Consequently, these our results warn that the deterministic description of 
high-temperature plasmas cannot capture important information of the 
turbulent systems and the statistical analyses with the 
Langevin equations and the PDFs are inevitably needed.

% If in two-column mode, this environment will change to single-column
% format so that long equations can be displayed. Use
%\begin{widetext}
%\end{widetext}

% entire page. There is no need to do explicit centering.

% \begin{figure}
% \end{figure}

% Surround figure environment with turnpage environment for landscape
% figure
% \begin{figure}
% \end{figure}

%\appendix
%\section{}

\begin{acknowledgments}
Nice discussions and critical reading of the manuscript by Prof.\ Yagi are 
acknowledged.
This work was supported by the Grant-in-Aid for Scientific Research of 
Ministry of Education, Culture, Sports, Science and Technology, 
the collaboration programmes of RIAM of Kyushu University and 
the collaboration programmes of NIFS.
\end{acknowledgments}

\bibliography{text}

\end{document}